\documentclass[12pt]{article}

\usepackage[utf8]{inputenc} 
\usepackage[T1]{fontenc} 
\usepackage[english]{babel}

\newenvironment{myframe}[1]{%
  \begin{framed}\noindent\textbf{#1}}{%
  \end{framed}%
}
\usepackage{amsmath,amssymb,amsfonts}
\usepackage{mathtools}
\usepackage{amsthm}

\usepackage{float}
\usepackage{moreverb,url}
\usepackage{hyperref}
\hypersetup{
    colorlinks,
    bookmarksopen,
    bookmarksnumbered,
    citecolor=blue,
    urlcolor=blue
}

\usepackage{framed}
\usepackage{authblk} 
\usepackage{booktabs}
\usepackage{multirow}
\usepackage{enumitem}

\newtheorem{remark}{Remark}
\newtheorem{theorem}{Theorem}

\newtheorem{assumption}{Assumption}

\title{Screening for Diabetes Mellitus in the U.S. Population Using Neural Network Models and Complex Survey Designs}
\author[1]{Marcos Matabuena}
\author[2]{Juan C. Vidal}
\author[3]{Rahul Ghosal}
\author[1]{Jukka-Pekka Onnela}
\affil[1]{Department of Biostatistics, Harvard University, USA}
\affil[2]{Research Center on Intelligent Technologies, University of Santiago de Compostela, SPAIN}
\affil[3]{Department of Epidemiology and Biostatistics, University of South Carolina, USA}
\date{} 

\begin{document}
\maketitle
\begin{abstract}
Complex survey designs are commonly employed in many medical cohorts. In such scenarios, developing case-specific predictive risk score models that reflect the unique characteristics of the study design is essential for minimizing selective biases in the statistical results. The objectives of this paper are to:
(i) propose a general predictive framework for regression and classification using neural network (NN) modeling that incorporates survey weights into the estimation process;
(ii) introduce an uncertainty quantification algorithm for model prediction tailored to data from complex survey designs; and
(iii) apply this method to develop robust risk score models for assessing the risk of Diabetes Mellitus in the US population, utilizing data from the NHANES 2011-2014 cohort.
The results indicate that models of varying complexity, each utilizing a different set of variables, demonstrate different discriminative power for predicting diabetes (with different economic cost), yet yield generalizable results at the population level.
Although the focus is on diabetes, this NN predictive framework is adaptable for developing clinical models across a diverse range of diseases and medical cohorts. The software and data used in this paper are publicly available on GitHub.
\end{abstract}

\section{Introduction}

Scientific experimental design is a key factor contributing to the biomedical reproducibility crisis, as highlighted by several studies \cite{begley2015reproducibility,an2018crisis,munafo2022reproducibility}. Although some practitioners emphasize the importance of study design, its influence is often underestimated. The significance of robust experimental design dates back to Ronald Fisher’s pioneering work in statistics \cite{yates1964sir} and remains a central topic in modern statistical science, particularly in the context of dynamic and adaptive designs. In medical research, a sound experimental design is essential for developing and validating new drugs and clinical treatments, especially within randomized clinical trials \cite{robertson2023response}.

Moreover, survey sampling methods used in nationally representative studies—such as the National Health and Nutrition Examination Survey (NHANES)—greatly influence the generalizability of clinical predictive models \cite{matabuena2022physical,matabuena2023distributional}. NHANES, renowned for its reliability and comprehensive data collection, employs a multi-stage complex sampling design to select a representative sample of the U.S. civilian non-institutionalized population. This approach incorporates hierarchical selection stages—from states to cities—and applies post-stratification techniques to mitigate non-response bias and improve population representativeness, thereby enhancing the efficiency of statistical estimators.

In contrast, other clinical datasets, such as the UK Biobank, often suffer from selection bias due to the voluntary nature of participant recruitment \cite{swanson2012uk,bradley2022addressing}. NHANES’s sophisticated experimental protocol, however, supports more reliable conclusions and serves as a benchmark for monitoring health behaviors in the U.S. population. The success of statistical analyses in such studies heavily depends on adopting methods that appropriately account for the underlying sampling design.

Developing risk assessment scores is critical for healthcare planning, particularly in public health, as these scores estimate the likelihood of disease onset and help identify individuals at elevated risk \cite{martin2009developing,hoeyer2019data,wiemken2020machine,dahlof2010cardiovascular}. Based on these assessments, healthcare strategies—such as tailored follow-ups, routine check-ups, and non-invasive interventions—can be implemented to reduce healthcare costs and improve population health outcomes \cite{hammouri2023uncertainty,turnbull2024population,parker2024economic}.

However, constructing disease-specific risk scores using observational data is inherently vulnerable to selection bias, which limits the generalizability of findings \cite{bradley2022addressing}. This issue is especially prominent in diabetes research, where the predictive performance of risk scores can vary substantially due to differences in genetic background, demographic factors, and study design quality. Practical limitations, including high costs and technical constraints, often hinder the implementation of efficient random sampling designs.

To overcome these challenges, researchers frequently rely on multi-stage survey designs, as exemplified by NHANES. These designs involve sampling across various hierarchical levels and targeting distinct subpopulations. Despite their widespread use, the integration of such survey designs with machine learning models and survey-weighted inference remains limited. Furthermore, the inferential properties of these models continue to pose significant challenges.

To bridge this gap and potentially enhance clinical conclusions in biomedical research (see Table~\ref{tab:my-table} for the NHANES case), we propose a novel neural network framework for prediction that incorporates uncertainty quantification through conformal prediction techniques. We apply this methodology to develop reliable risk scores for detecting Diabetes Mellitus in the U.S. population.

\begin{table*}[t!] 
\small\sf\centering 
  \caption{Impacts of not using survey weights in NHANES data analysis} 
\begin{tabular}{@{}ll@{}} 
\toprule 
\textbf{Aspect} & \textbf{Impact of Not Using Correct NHANES Survey Weights} \\ 
\midrule 
Bias in estimates & Biased statistical results, misrepresenting certain groups \\ 
Loss of representativeness & Results not reflective of the U.S. civilian non-institutionalized population \\ 
Policy decisions & Potential misinformed public health policies and resource allocation \\ 
Statistical inaccuracy & Incorrect standard errors, confidence intervals, and p-values \\ 
Ethical considerations & Ethical concerns due to disproportionate exclusion of minority groups \\ 
\bottomrule 
\end{tabular} 
\label{tab:my-table} 
\end{table*}

\subsection{Diabetes study case}

Diabetes Mellitus represents a significant public health challenge, currently affecting approximately 12\% of the U.S. population. A notable concern, particularly in Type 2 Diabetes, is the high rate of undiagnosed cases. The CDC in 2020 reported that around 21\% of diabetes cases in the U.S. remained undetected \cite{cdc2020national}. Our recent research suggests that this percentage could be even higher \cite{heilmann2023precise}.

The prevalence of sedentary lifestyles, especially in developed countries, contributes to a worrying projected increase in diabetes cases. Currently affecting about 9.3\% of the global population, projections suggest an increase to 10.2\% by 2030 and 10.9\% by 2045. This rising trend underscores the urgent need for effective public health strategies.

The economic impact of diabetes is substantial, with the total cost in the U.S. estimated at \$412.9 billion, including \$306.6 billion in direct medical expenses and \$106.3 billion in indirect costs \cite{parker2024economic}. The burden of diabetes extends beyond financial costs, significantly affecting life expectancy and quality of life, primarily due to late diagnoses and poor glycemic control. These issues often lead to severe complications, including cardiovascular problems. There is an increasing need to develop new precision medicine strategies based on a data-driven approach. These strategies can involve the development of new screening methods and the prescription of treatments with dynamic information obtained from wearable devices and electronic records.

In the case of Diabetes Mellitus, the diagnosis typically involves biomarkers such as glycosylated hemoglobin (A1C) and fasting plasma glucose (FPG). While FPG tests are cost-effective, A1C testing, a primary biomarker for diabetes, is more expensive and complex. Despite its widespread use, particularly in non-risk groups, A1C testing presents significant challenges. Medical guidelines, including those from the American Diabetes Association (ADA), recommend using both A1C and FPG, along with oral glucose tolerance tests, especially in cases of gestational diabetes \cite{kotzaeridi2021characteristics},  for a comprehensive assessment.

In disease diagnosis and screening, statistical and machine learning models, leveraging clinical variables, show promise as a tool in the identification of high-risk individuals. These predictive models can stratify patients effectively, allowing for personalized clinical approaches in line with precision medicine. However, their predictive power can vary across different patient demographics and in real-world applications with limited sample sizes, which poses challenges in accurately determining patients' glucose status.

In this paper, we address these challenges by utilizing data from the National Health and Nutrition Examination Survey (NHANES) 2011-2014. Given the complexity of the survey design and the relative scarcity of non-parametric models in this domain, we introduce novel neural network estimators. These estimators are designed to ensure universal approximation capabilities in predictive tasks. Furthermore, we developed a new uncertainty quantification framework based on conformal inference techniques for survey data, enhancing our ability to quantify the predictive limits of these models. This holistic approach aims to improve the precision and applicability of predictive models for Diabetes Mellitus.

\subsection{Neural network models and machine learning models for survey data}

The application of non-parametric regression models in survey data analysis remains a relatively unexplored area \cite{lumley2017fitting}. Traditional approaches, such as the Nadaraya-Watson estimator \cite{harms2010kernel} and local polynomial regression, have been proposed, but they tend to be highly sensitive to a large number of predictors. Recently, machine learning algorithms—such as neural networks and random forests—have been recommended in the literature as non-parametric alternatives to classical regression models. However, their generalization to inferential and predictive tasks in complex survey designs has not yet been adequately addressed. To the best of our knowledge, the closest related work is the kernel ridge regression models proposed in \cite{matabuena2023distributional} for functional and distributional data analysis applications \cite{matabuena2022physical}.

The goal of this paper is to address this gap in the literature by proposing a general neural network (NN) framework for classification and regression models, equipped with automated uncertainty quantification for model predictions.

Given the common prevalence of nonlinear biological data, the universal approximation properties of neural networks—combined with their robustness in high-dimensional settings—make them excellent candidates for modeling tasks across various domains. In particular, for biomedical applications, the proposed NN predictive framework offers a novel opportunity to develop disease-specific risk scores that are both accurate and trustworthy. This is largely due to the incorporation of uncertainty quantification directly into the predictive outputs of the models.

\section*{Summary of contributions}

The contributions of this article are summarized below:

\begin{itemize}
    \item \textbf{Neural network models for survey data:} We introduce a novel NN predictive modeling strategy for complex survey data, with a particular focus on developing disease risk scores within cohorts such as NHANES.
    \item \textbf{Prediction uncertainty quantification:} We propose a strategy for quantifying prediction uncertainty using conformal inference techniques adapted for survey designs. While assumptions such as exchangeability are often violated in complex survey settings, and strict non-asymptotic guarantees may not hold, we demonstrate that our algorithms remain consistent in the asymptotic regime.
     
    \item \textbf{Versatile framework:} Our NN predictive framework supports both regression and classification tasks. In particular, we propose a new neural network-based quantile regression algorithm to enable survey-based conformal prediction. This contribution fills a notable gap in the literature on quantile regression using non-linear models in the context of survey data \cite{10.1093/biostatistics/kxab035}.
    
    \item \textbf{Diabetes application analysis:} We evaluate models of varying complexity and economic cost in the context of diabetes detection. Our analysis highlights trade-offs between predictive performance and implementation cost across subpopulations.
    
    \item \textbf{Software for the scientific community:} To ensure reproducibility and facilitate adoption, the Python software implementing our proposed methods is publicly available at XXX.
\end{itemize}

\section*{Outline of the paper}

The structure of the paper is as follows. Section~\ref{sec:models} introduces our neural network models designed specifically for survey data, along with the proposed approach for uncertainty quantification. In Section~\ref{sec:sim}, we present a simulation study in a binary classification setting. Section~\ref{sec:NHANES} applies the models to a diabetes case study using NHANES data. Section~\ref{sec:discussion} discusses the methodological contributions, broader applications in biomedical research, and the practical implications of our findings.

\section{Methodology}\label{sec:models}

\subsection{Model estimation: Survey neural network models}

Suppose we observe a random sample $\mathcal{D}_{n} = \{(X_i, Y_i) \sim P\}_{i=1}^{n}$ drawn from a finite population $\mathcal{M}$ of size $N$, according to a complex survey design. For each unit $i$ $(i = 1, \dots, n)$, we associate a weight $w_i$ that reflects the number of population units represented by the $i$-th sampled unit. We assume that $w_i = 1/\pi_i$, where $\pi_i$ denotes the probability of selecting unit $i$ under the sampling design. In our scientific application, the final weights $w_i$ incorporate post-stratification corrections to account for nonresponse, following the NHANES survey design guidelines.\footnote{\url{https://www.cdc.gov/nchs/nhanes/tutorials/weighting.aspx}}

Without loss of generality, we assume that each predictor $X_i \in \mathbb{R}^{p}$, and the response variable $Y_i$ either belongs to a set of categories $\{1, \dots, K\}$ in a multi-class classification setting or is a scalar continuous variable $Y_i \in \mathbb{R}$ in the case of regression. For practical purposes, the modeling framework remains unchanged, with the only difference being the choice of the loss function $\ell$ appropriate to the outcome type.

Deep neural networks (DNNs) \cite{fan2021selective} model nonlinear relationships using compositions of nonlinear transformations. Formally, we define

\begin{equation*}
h^{(L)} = g^{(L)} \circ g^{(L-1)} \circ \dots \circ g^{(1)},
\end{equation*}

\noindent where $\circ$ denotes function composition, $L$ is the number of hidden layers, and $h^{(0)} = x$. Recursively, each hidden layer is defined as $h^{(l)} = g^{(l)}(h^{(l-1)})$ for $l = 1, \dots, L$. In feed-forward neural networks, we specify

\begin{equation}
h^{(l)} = \sigma\left(\omega^{(l)} h^{(l-1)} + b^{(l)}\right),
\end{equation}

\noindent where $\omega^{(l)}$ is the weight matrix and $b^{(l)}$ is the bias vector at layer $l$, for $l = 1, \dots, L$. The activation function $\sigma(\cdot)$ is applied element-wise; in this work, we use the ReLU function:

\begin{equation}
\sigma(z_j) = \max(z_j, 0),
\end{equation}

\noindent where $z_j$ denotes the $j$-th coordinate of the vector $z$.

Given the final hidden layer output $h^{(L)}$ and observed label $Y_i$, we estimate the model parameters by minimizing a weighted loss function. Specifically, we adopt a weighted cross-entropy loss to account for survey weights. The set of parameters $\theta$ is obtained by solving the following optimization problem:

\begin{equation}
\begin{aligned}
\widehat{\theta} = \arg\min_{\theta} \sum_{i=1}^{n} w_i \cdot \ell(Y_i, \log f(X_i; \theta)) =
\\ \arg\min_{\theta} \sum_{i=1}^{n} \sum_{k=1}^{K} \mathbb{I}\{Y_i = k\} \cdot w_i \cdot \log f_k(X_i; \theta),
\end{aligned}
\end{equation}

\noindent where $\theta = \{\omega^{(l)}, b^{(l)} \mid 1 \leq l \leq L+1\}$ denotes the model parameters. The function $f_k(x; \theta)$ represents the $k$-th component of the softmax output:

\begin{equation}
f_k(x; \theta) = \frac{\exp(z_k)}{\sum_{j=1}^{K} \exp(z_j)}, \quad \forall k \in \{1, \dots, K\},
\end{equation}

\noindent where $z = \sigma(\omega^{(L+1)} h^{(L)} + b^{(L+1)}) \in \mathbb{R}^{K}$.

In the context of complex survey data, estimators that incorporate sampling weights—such as those based on the Horvitz-Thompson estimator—are commonly used to produce design-consistent and unbiased estimates. These approaches adjust for unequal probabilities of selection, improving efficiency and representativeness. Our method follows this tradition by integrating the survey weights into the neural network training procedure through the loss function.

\begin{figure}[ht!]
	\centering
	\includegraphics[width=0.8\linewidth]{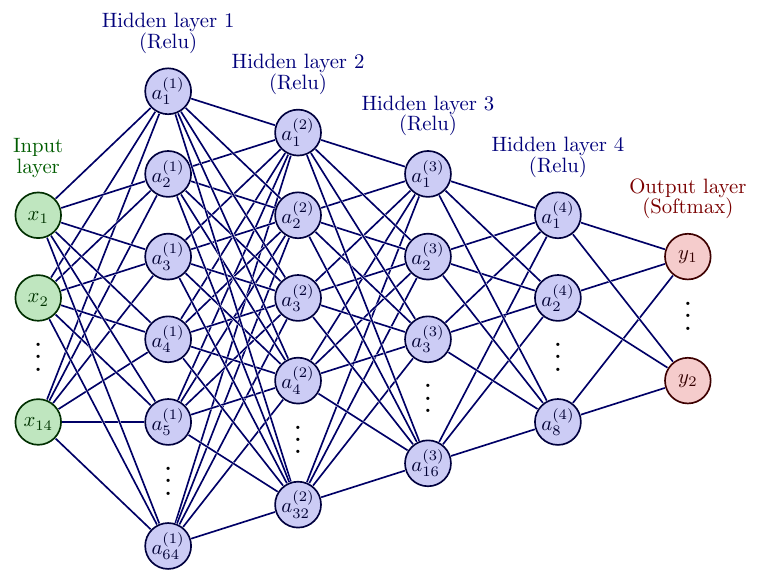} 
	\caption{Architecture of the neural network model used in this study.}
	\label{fig:graph24}
\end{figure}

\subsection{Computational details of NN}

In our scientific application, we focus on a binary classification problem, distinguishing between two classes: i) diabetes and ii) non-diabetes. Our approach utilizes a neural network architecture with a depth of \(L\) and two layers at each level. An abstract representation of the neural network architecture can be found in Figure~\ref{fig:graph24}. To fine-tune the model and select the optimal architecture, we partition the dataset \(\mathcal{D}_n\) into three disjoint subsets: training (\(\mathcal{D}_{\text{train}}\)), architecture selection (\(\mathcal{D}_{\text{architecture}}\)), and testing (\(\mathcal{D}_{\text{test}}\)), with proportions of 50\%, 30\%, and 20\% respectively. We adopt a cross-validation strategy, repeating the model evaluation process $10$ times to ensure robust validation of the model's performance across various metrics.

For the subject $i$-th, we denote the  predicted label as \(\widetilde{Y}_i\) and the estimated conditional probability of having diabetes as \(\widetilde{p}_{i}\approx 
\mathbb{P}(Y_i=1\mid X_i)\). We define \(D\) as the set of indices for diabetes patients and \(\overline{D}\) for non-diabetes patients. The performance metrics considered in the model evaluation are detailed in Table~\ref{tab:performance_metrics}.

To optimize the model parameters, we employ stochastic gradient descent (SGD) with the Adam algorithm, leveraging its adaptive learning rate properties. This optimization is facilitated by the auto-differentiation capabilities of the PyTorch computational framework.

\begin{table*}[!ht]
\centering
\caption{Performance metrics for binary classification with survey weights}
\label{tab:performance_metrics}

\scalebox{0.9}{%
\begin{tabular}{lcl}
\toprule
\textbf{Metric} & \textbf{Formula} & \textbf{Description} \\
\midrule
AUC 
  & $\displaystyle
     \sum_{i\in D} \sum_{j\in \overline{D}} w_i w_j
       \Bigl(\frac{1}{\sum_{i\in D} \sum_{j\in \overline{D}} w_i w_j}\Bigr)
       \,k(\widetilde{p}_i, \widetilde{p}_j)$ 
  & Weighted classification performance measure \\

Accuracy 
  & $\displaystyle \frac{TP+TN}{TP+TN+FP+FN}$ 
  & Proportion of correctly classified instances \\

Recall 
  & $\displaystyle \frac{TP}{TP+FN}$ 
  & Sensitivity (true positive rate) \\

Precision 
  & $\displaystyle \frac{TP}{TP+FP}$ 
  & Positive predictive value \\

F\textsubscript{1}-score 
  & $\displaystyle \frac{2 \times \text{PREC} \times \text{REC}}{\text{PREC}+\text{REC}}$ 
  & Harmonic mean of precision and recall \\

Cross-Entropy 
  & $\displaystyle -\sum_{i} w_i \Bigl[ y_i \log(\widetilde{p}_i) 
    + (1 - y_i) \log\bigl(1 - \widetilde{p}_i\bigr) \Bigr]$
  & Weighted log loss between predicted and actual labels \\
\bottomrule
\end{tabular}
}
\end{table*}

\subsection{Uncertainty quantification for survey data via conformal prediction}

\subsubsection{Background on conformal prediction and uncertainty quantification}. 

In medical studies, considerable uncertainty is common. Patient responses often vary over time and can exhibit individualized patterns \cite{hammouri2023uncertainty,banerji2023clinical}. Although uncertainty is frequently viewed negatively, it can provide valuable insights in clinical decision-making \cite{hammouri2023uncertainty}.

First, it helps clarify the limitations of predictive models. Second, high uncertainty in patient outcomes may stimulate the development of novel pharmacological treatments or interventions. Third, uncertainty supports healthcare planning—patients with unpredictable clinical courses may require more frequent follow-up. Finally, estimating individual-level uncertainty is crucial for identifying atypical cases and for allocating healthcare resources. Each of these factors underscores the value of uncertainty quantification in clinical practice.

Let $(X_{1},Y_{1}), \ldots, (X_{n}, Y_{n})$ be a random sample that is at least exchangeable. We aim to extend the conformal prediction framework to derive uncertainty measures for regression models with scalar responses.

Consider the sequence $\mathcal{D}_{n} = \{(X_i, Y_{i})\}^{n}_{i=1}$ of i.i.d. random variables. Given a new i.i.d. random pair $(X_{n+1}, Y_{n+1})$, conformal prediction, as introduced by \cite{vovk2005algorithmic}, provides a flexible family of algorithms for constructing prediction intervals that are valid regardless of the underlying regression method.

Let us consider a regression algorithm
\[
\mathcal{A}: \bigcup_{n\geq 0} (\mathcal{X}\times \mathbb{R})^n \rightarrow 
\{\textnormal{measurable function } \widetilde{m}: \mathcal{X}\rightarrow\mathbb{R}\}, 
\]
which maps any training dataset to a fitted regression function $\widetilde{m}$. The algorithm $\mathcal{A}$ must treat the data points symmetrically, that is,
\begin{equation}\label{eqn:alg_symmetric}
\begin{aligned}
\mathcal{A}\big((x_{\pi(1)},y_{\pi(1)}),\dots,(x_{\pi(n)},y_{\pi(n)})\big) = 
\mathcal{A}\big((x_1,y_1),\dots,(x_n,y_n)\big)
\end{aligned}
\end{equation}
for all $n \geq 1$, all permutations $\pi$ on $[n]=\{1,\dots,n\}$, and all data sets $\{(x_i,y_i)\}_{i=1}^{n}$. For each $y \in \mathbb{R}$, let
\[
\widetilde{m}^{y} = \mathcal{A}\big((X_1,Y_1),\dots,(X_n,Y_n),(X_{n+1},y)\big)
\]
denote the regression model fitted with the additional point $(X_{n+1},y)$. Define the conformity scores:
\begin{equation}\label{eqn:R_y_i}
R^y_i = 
\begin{cases}
|Y_i - \widetilde{m}^{y}(X_i)|, & i=1,\dots,n,\\ 
|y - \widetilde{m}^{y}(X_{n+1})|, & i=n+1.
\end{cases}
\end{equation}
Then the conformal prediction interval for $X_{n+1}$ is
\begin{equation}\label{eqn:def_fullCP}
\begin{aligned}
\widetilde{C}^{\alpha}(X_{n+1};\mathcal{D}_{n}) = 
\Big\{ y \in \mathbb{R} \ : \ R^y_{n+1} \leq 
\textnormal{quant}_{1-\alpha}\Big(
\sum_{i=1}^{n+1} \frac{1}{n+1} \cdot \delta_{R^y_i}
\Big) \Big\},
\end{aligned}
\end{equation}
where $\textnormal{quant}_{1-\alpha}(\cdot)$ denotes the $(1-\alpha)$ quantile of the empirical distribution.

The full conformal method provides distribution-free coverage guarantees:

\begin{theorem}[Full Conformal Prediction \cite{vovk2005algorithmic}]\label{thm:background_fullCP}
If the data points $(X_1,Y_1),\dots,(X_n,Y_n),(X,Y)$ are i.i.d.\ (or more generally, exchangeable), and the algorithm treats data symmetrically as in~\eqref{eqn:alg_symmetric}, then the conformal prediction set~\eqref{eqn:def_fullCP} satisfies
\[
\mathbb{P}(Y_{n+1}\in \widetilde{C}^{\alpha}(X_{n+1}; \mathcal{D}_{n})) \geq 1-\alpha.
\]
The same guarantee holds for the split conformal method, which separates training and calibration data.
\end{theorem}

Conformal inference (both split and full) is often described in terms of nonconformity scores $\widetilde{S}(X_i,Y_i)$, which measure how atypical a data point is with respect to a fitted model. The standard score is the residual:
\begin{equation}\label{eqn:standard_nonconformity_score}
\widetilde{S}(X_i,Y_i) := |Y_i - \widetilde{m}(X_i)|,
\end{equation}
where $\widetilde{m}$ is fitted using a subset of the training data.

\subsubsection{NHANES study design}.

The NHANES cohorts use a sophisticated multistage probability sampling design to represent the U.S. civilian non-institutionalized population. Sampling begins with Primary Sampling Units (PSUs), usually counties or groups of counties, selected with probability proportional to size. PSUs are subdivided, households are sampled, and individuals are chosen using stratification criteria such as age, sex, and race/ethnicity.

However, demographic and geographic heterogeneity across states—differences in ethnic makeup, socioeconomic status, and health outcomes—may violate the exchangeability assumption. This poses challenges for applying conformal prediction algorithms that rely on exchangeability or i.i.d.\ assumptions for finite-sample validity.

\subsubsection{Conformal prediction under covariate shift for survey data}. 

To accommodate survey settings like NHANES, we focus on scenarios in which the data $(X_i, Y_i)$ for $i=1,\ldots,n+1$ are no longer exchangeable. Instead, we assume a covariate shift setting:
\begin{gather*}
(X_{i}, Y_{i}) \stackrel{\text{i.i.d.}}{\sim} P = P_{X} \times P_{Y|X}, \hspace{0.2cm} i=1, \ldots, n, \\
(X_{n+1}, Y_{n+1}) \sim \widetilde{P} = \widetilde{P}_{X} \times P_{Y|X},
\end{gather*}
i.e., the conditional distribution $P_{Y|X}$ remains unchanged, but the marginal $P_X$ may shift to $\widetilde{P}_X$.

Assuming the likelihood ratio $w(x) = \frac{d\widetilde{P}_X}{dP_X}(x)$ is known or estimated, we define a weighted conformal prediction method using the distribution:
\[
\sum_{i=1}^{n} p_{i}^{w}(x) \delta_{V_{i}^{(x, y)}} + p_{n+1}^{w}(x) \delta_{\infty},
\]
with weights
\[
p_{i}^{w}(x) = \frac{w(X_{i})}{\sum_{j=1}^{n} w(X_{j}) + w(x)}, \quad 
p_{n+1}^{w}(x) = \frac{w(x)}{\sum_{j=1}^{n} w(X_{j}) + w(x)}.
\]

\begin{theorem}
Under the covariate shift model above, assume $\widetilde{P}_X \ll P_X$ and define the weighted predictive set:
\begin{equation}
\begin{aligned}
\widetilde{C}_{n}(x) = \Big\{ y \in \mathbb{R} : V_{n+1}^{(x, y)} \leq 
\operatorname{Quantile}\Big(1-\alpha ; \\
\sum_{i=1}^{n} p_{i}^{w}(x) \delta_{V_{i}^{(x, y)}} + p_{n+1}^{w}(x) \delta_{\infty} \Big) \Big\},
\end{aligned}
\end{equation}
where $V_i^{(x,y)}$ are residuals or nonconformity scores. Then
\[
\mathbb{P}\left\{Y_{n+1} \in \widetilde{C}_{n}(X_{n+1})\right\} \geq 1-\alpha.
\]
\end{theorem}

For classification, we adapt the quantile classification algorithm of \cite{cauchois2021knowing} to handle survey weights and covariate drift.

\begin{myframe}{Split conformalized quantile classification (CQC) under survey design}

\vspace{2mm}

\noindent \textbf{Input:} Sample $\{\left(X_i,Y_i\right)\}_{i=1}^{n},$ index sets $\mathcal{I}_{1}, \mathcal{I}_{2}, \mathcal{I}_{3},$ algorithm $\mathcal{A},$ quantile function class $\mathcal{Q},$ confidence level $\alpha$.

\begin{enumerate}
    \item Compute fitted score $\widetilde{s}_i = \sum_{k=1}^{K} 1\{Y_i=k\} w_i \log f_{k}(X_i;\theta)$ for $i \in \mathcal{I}_1$.
    
    \item Fit the quantile regression model:
    \[
    \widetilde{q}_{\alpha} = \arg \min_{q \in \mathcal{Q}} \frac{1}{|\mathcal{I}_{2}|} \sum_{i \in \mathcal{I}_{2}}  w_i \rho_{\alpha}(\hat{s}_i - q(X_i)),
    \]
    where $\rho_{\alpha}(t) = (1-\alpha)[-t] + \alpha[t]$.
    
    \item Calibrate using nonconformity scores $S_i = \widetilde{q}_{\alpha}(X_i) - \widetilde{s}_i$. Define the weighting empirical quantile:
    \[
    Q_{1-\alpha}(S,\mathcal{I}_{3}) := \left[1+\frac{1}{n_3}\right] (1-\alpha)\text{-quantile of } \{S_i\}_{i \in \mathcal{I}_3},
    \]
    and return the predictive set:
    \[
    \widetilde{C}_{n,1-\alpha}(x) := \left\{k \in \{1,\dots,K\} \ \middle| \ \widetilde{s}(x,k) \geq \widetilde{q}_{\alpha}(x) - Q_{1-\alpha}(S,\mathcal{I}_{3}) \right\}.
    \]
\end{enumerate}
\label{alg:conformal}
\end{myframe}

\subsubsection{Conformal prediction beyond exchangeability}.

More generally, \cite{barber2022conformal} introduced conformal methods for non-exchangeable data. A key concept is the \textit{coverage gap}, which quantifies the deviation from the nominal level:
\[
\text{Coverage gap} = (1-\alpha) - \mathbb{P}\left\{Y_{n+1} \in \widetilde{C}_{n}(X_{n+1})\right\}.
\]

Let $Z = (Z_1, \ldots, Z_{n+1})$ denote the full data and $Z^i$ the same data with the $i$th point swapped with $(X_{n+1}, Y_{n+1})$. Using weights $w_i \in [0,1]$ assigned to each point, the coverage gap can be bounded by:
\[
\text{Coverage gap} \leq \frac{\sum_{i=1}^{n} w_i \cdot \mathrm{d}_{\mathrm{TV}}(Z, Z^i)}{1 + \sum_{i=1}^{n} w_i},
\]
where $\mathrm{d}_{\mathrm{TV}}$ is the total variation distance.

In practice, this gap reflects the extent to which survey sampling violates exchangeability. In our NHANES case study, we evaluate the empirical coverage gap and demonstrate that, under asymptotic consistency, the gap vanishes as $n \to \infty$.

\section{Simulation study}\label{sec:sim}
We conducted a simulation study employing a two-stage cluster sampling design, based on two distinct generative models for binary regression, where \( Y \in \{0,1\} \). In both scenarios, the predictor variable is defined as \( X = (X_1, \dots, X_5)^{\top} \in \mathbb{R}^5 \), with \( X \sim \mathcal{N}_5(\mu, \Sigma) \), where \( \mathcal{N}_5(\mu, \Sigma) \) denotes a 5-dimensional multivariate normal distribution with mean vector \( \mu \) and covariance matrix \( \Sigma \). For simplicity, we set \( \mu = \mathbf{0}_{5\times 1} \), a five-dimensional zero vector, and \( \Sigma = \mathcal{I}_{5\times 5} \), the \( 5 \times 5 \) identity matrix.

The conditional distribution of the response variable is given by \( Y \mid X \sim \text{Ber}(\pi(X)) \), where \( \pi(X) = \mathbb{P}(Y = 1 \mid X) \in [0, 1] \) is the success probability, specified through the following two functional forms:
\begin{itemize}
    \item[(a)] \( \pi(X) = \frac{1}{1 + \exp\left(-3 + \sum_{i=1}^{3} X_i\right)} \)
    \item[(b)] \( \pi(X) = \frac{1}{1 + \exp\left(-2 + \sum_{i=1}^{3} X_i + X_4 X_5\right)} \)
\end{itemize}

The experimental design incorporates assumptions regarding the hierarchical structure of clusters. In the first-stage design, clusters represent states and regions, assuming independence of selection probabilities across these clusters. In the second stage, clusters represent cities, again assuming independence in the selection probabilities for each city.

Each state is assumed to have a selection probability of $4/5$, and within each state $i$, there exists a variable number of cities, denoted by $n_i$. Each city within a given state is selected with equal probability $1/n_{i}$. The sampling mechanism and the computation of individual patient selection probabilities ($w_{r}$) are implemented using the \texttt{survey} package in \textbf{R}.

For each generative model (a and b), we replicated the experimental design 500 times ($b=1,\dots,B=500$) across varying sample sizes $n \in \{5000, 10000, 20000\}$. The performance was evaluated using several metrics: (i) AUC, (ii) accuracy, (iii) recall, (iv) precision, (v) F1 score, and (vi) cross-entropy. Comparisons were drawn between a neural network (NN) approach and a competing method, specifically a logistic regression model with survey weights.

Results, reported as mean $\pm$ standard deviation across the simulations, are summarized in Table~\ref{table:sim}. As expected, in scenario (a), where the underlying logistic regression model represents the ground truth, the accuracy of the NN algorithm and the parametric logistic regression model are similar, particularly at larger sample sizes. However, in scenario (b), characterized by nonlinear interactions, the NN algorithm consistently outperforms the logistic model.

Consistency across performance metrics is observed for fixed sample sizes. Generally, increasing the sample size $n$ improves accuracy, reduces errors, and yields more stable results with lower variability.

\begin{table}[tb!]
\caption{Simulation results for GLM and MLP under scenarios (a) and (b), with $N \in \{5000, 10000, 20000\}$. Mean ± SD for: AUC, accuracy, recall, precision, F1, and binary cross-entropy.}  
\centering
\scriptsize
\setlength{\tabcolsep}{3pt}
\renewcommand{\arraystretch}{0.9}
\begin{tabular}{rlllllll}
  \hline
  & & \multicolumn{3}{c}{a)} & \multicolumn{3}{c}{b)} \\
  \cline{3-5} \cline{6-8}
  & & 5k & 10k & 20k & 5k & 10k & 20k \\ 
  \hline
  \multirow{6}{*}{\rotatebox{90}{GLM}}
  & AUC      & 0.9325±0.0182 & 0.9328±0.0161 & 0.9332±0.0142 & 0.9998±0.0011 & 0.9998±0.0008 & 0.9999±0.0007 \\ 
  & Acc.     & 0.9039±0.0178 & 0.9041±0.0153 & 0.9048±0.0135 & 0.9980±0.0030 & 0.9985±0.0024 & 0.9989±0.0021 \\ 
  & Prec.    & 0.8438±0.0446 & 0.8485±0.0390 & 0.8506±0.0344 & 0.9968±0.0071 & 0.9976±0.0057 & 0.9982±0.0048 \\ 
  & Recall   & 0.7660±0.0530 & 0.7675±0.0457 & 0.7692±0.0405 & 0.9970±0.0065 & 0.9978±0.0051 & 0.9983±0.0043 \\ 
  & F1       & 0.8017±0.0381 & 0.8050±0.0328 & 0.8070±0.0290 & 0.9969±0.0048 & 0.9977±0.0039 & 0.9982±0.0033 \\ 
  & CE       & 0.2791±0.0368 & 0.2801±0.0325 & 0.2796±0.0288 & 0.0218±0.0373 & 0.0164±0.0301 & 0.0127±0.0257 \\ 
  \hline
  \multirow{6}{*}{\rotatebox{90}{MLP}}
  & AUC      & 0.9843±0.0096 & 0.9892±0.0087 & 0.9920±0.0081 & 0.9998±0.0003 & 0.9999±0.0003 & 0.9999±0.0002 \\ 
  & Acc.     & 0.9495±0.0160 & 0.9580±0.0157 & 0.9644±0.0159 & 0.9923±0.0061 & 0.9940±0.0050 & 0.9949±0.0045 \\ 
  & Prec.    & 0.9244±0.0370 & 0.9383±0.0336 & 0.9465±0.0312 & 0.9880±0.0135 & 0.9906±0.0113 & 0.9921±0.0099 \\ 
  & Recall   & 0.8741±0.0501 & 0.8969±0.0481 & 0.9878±0.0141 & 0.9878±0.0141 & 0.9904±0.0115 & 0.9920±0.0101 \\ 
  & F1       & 0.8975±0.0331 & 0.9163±0.0331 & 0.9294±0.0333 & 0.9878±0.0096 & 0.9904±0.0080 & 0.9920±0.0071 \\ 
  & CE       & 0.1423±0.0340 & 0.1185±0.0360 & 0.1014±0.0384 & 0.0212±0.0097 & 0.0166±0.0089 & 0.0137±0.0085 \\ 
  \hline
\end{tabular}
\label{table:sim}
\end{table}

\begin{figure}[H]
	\centering
	\includegraphics[width=1\linewidth,height=0.5\linewidth]{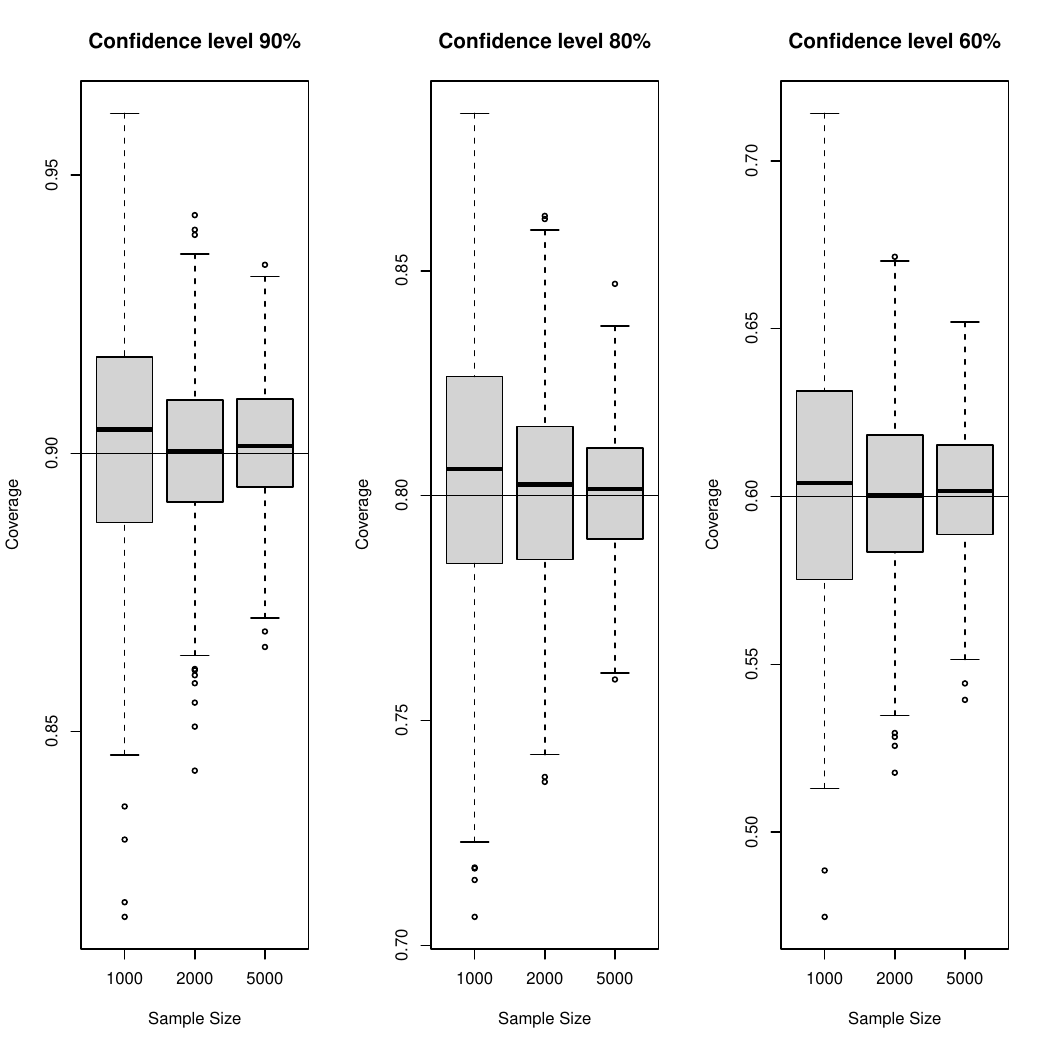} 
	\caption{Results of conformal simulation exercise for the generative model a)}
	\label{fig:conformal1}
\end{figure}

\subsection{Uncertainty quantification}

\begin{figure}[ht]
	\centering
	\includegraphics[width=0.99\linewidth,height=0.5\linewidth]{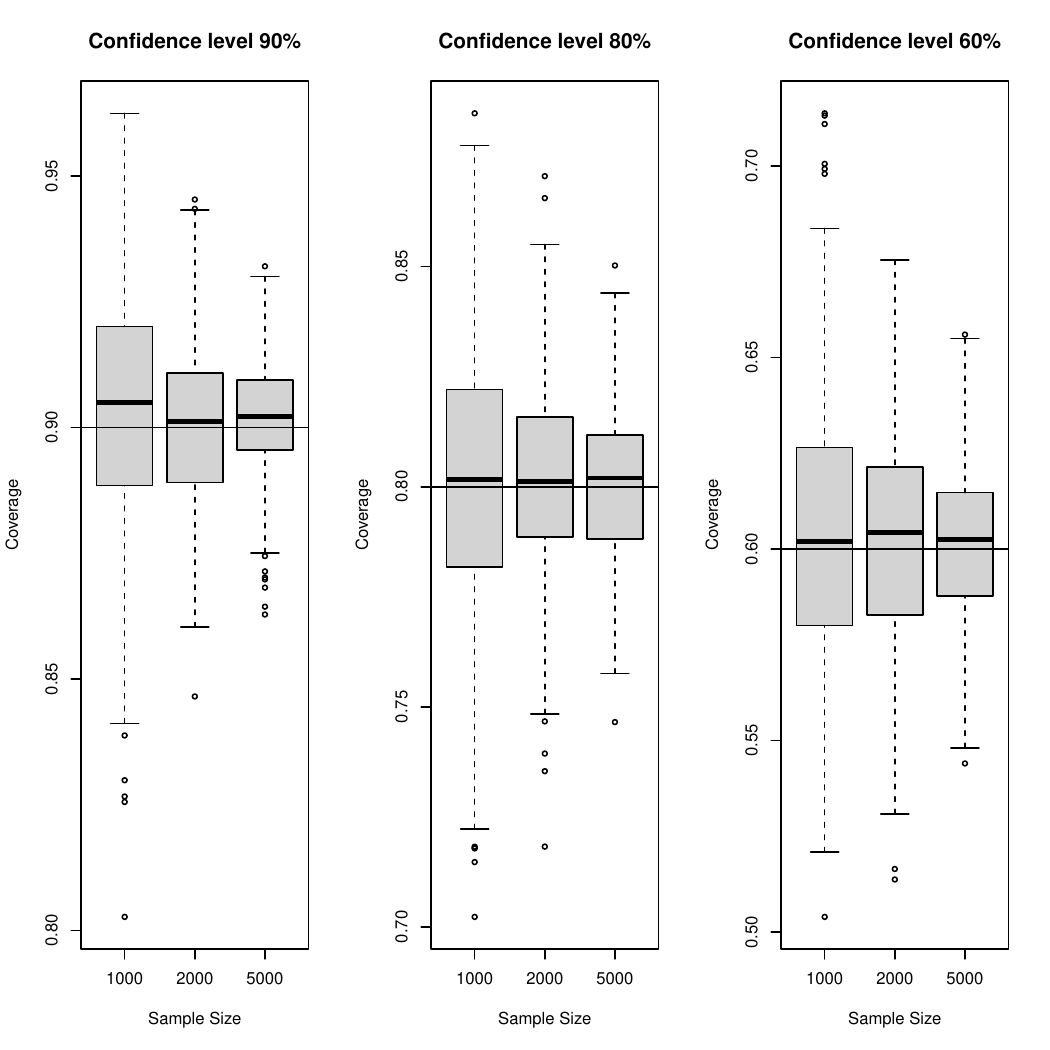} 
	\caption{Results of conformal simulation exercise for the generative model b)}
	\label{fig:conformal2}
\end{figure}

In the context of conformal prediction for survey data, we focus again on the generative models from the simulation scenarios (a) and (b). However, shift to evaluating model performance based on marginal coverage across different confidence levels $\alpha=0.95, 0.9, 0.8$. Specifically, we assess the independent random sample $\mathcal{D}_{\text{test}}$, comprising $N=5000$ data points, by quantifying the probability metrics $\mathbb{P}(Y\in \widetilde{C} ^{\alpha}(X))$ across multiple simulations $b=1,\dots, B$. The results of this simulation exercise for scenarios (a) and (b) are depicted in Figures \ref{fig:conformal1} and \ref{fig:conformal2}, respectively.

Our findings show the non-asymptotic property of conformal prediction, demonstrating that $\mathbb{P}(Y\in \widetilde{C}^{\alpha}(X))\geq 1-\alpha$. This observation aligns with our experimental design, which conditions data to be exchangeable across clusters. Notably, the boxplots illustrate a convergence toward the nominal value as the sample size $N$ increases. Furthermore, the variability in the boxplots diminishes, indicating consistent and reliable model performance in these specific simulation examples.

\section{NHANES diabetes case study}\label{sec:NHANES}

\subsection{Literature review: Predictive models for diabetes risk stratification}

Various predictive models have been developed to stratify diabetes risk, including the Finnish (FINDRISC) \cite{makrilakis2011validation} and German (GDRS) \cite{muhlenbruch2018derivation} diabetes scores, formulated over a decade ago. These scores utilize logistic regression to predict the 10-year risk of diabetes or employ survival analysis techniques, such as Cox regression, to estimate the time to diabetes onset. They rely on easily obtainable variables, such as age, sex, anthropometric measures, lifestyle, family medical history, and medication use.

Recent approaches have embraced machine learning (ML) methodologies to predict diabetes progression from healthy or prediabetic states, exhibiting promising results \cite{wu2021,cahn2020}. Nevertheless, these studies are primarily based on observational data, frequently excluding subjects with incomplete data, necessitating caution when applying these results clinically.

Our approach differs by focusing on predicting diabetes status at a specific point in time using current patient characteristics. This approach provides statistical advantages, increases case numbers, and mitigates imbalance issues common in longitudinal prediction models. Additionally, we incorporate a survey-sampling design, enabling robust, generalizable predictions for the U.S. population.

\subsection{NHANES 2011--2014 data}

We utilized data from the National Health and Nutrition Examination Survey (NHANES) waves 2011--2014 \cite{johnson2014national}, a comprehensive survey targeting the U.S. civilian, non-institutionalized population.

Data collection involved both interviews and clinical examinations. Interviews gathered demographic, health, and nutritional information, whereas clinical examinations included physical measurements, blood pressure assessment, dental exams, and collection of blood and urine samples for laboratory analysis. Our dataset comprises 5011 individuals aged 10 to 80 years.

Key variables considered in our study include age (categorical and continuous), race, gender, diagnoses of cancer or diabetes (categorical variables), blood pressure, grip strength, body mass index (BMI), and biochemical biomarkers such as cholesterol and triglycerides (continuous variables).

Race categories were encoded as: 1 = Mexican American; 2 = Other Hispanic; 3 = Non-Hispanic white; 4 = Non-Hispanic black; 5 = Non-Hispanic Asian; 6 = Other Race, including multi-racial. Table \ref{tab:describ} summarizes these variables.

\begin{table}[H]
\centering
\caption{Clinical characteristics of diabetes and non-diabetes patients}
{\footnotesize
\begin{tabular}{lll}
\hline
\textbf{Variable} & \textbf{Diabetes (32\%)} & \textbf{No Diabetes (68\%)} \\
\hline
Age (years) & $54.1 \pm 15.2$ & $44.8 \pm 16.2$ \\
Height (cm) & $88.3 \pm 22.5$ & $80.8 \pm 19.3$ \\
Weight (kg) & $169.1 \pm 10.1$ & $169.4 \pm 9.6$ \\
BMI (kg/m$^2$) & $30.8 \pm 7.13$ & $28.08 \pm 6.16$ \\
Waist circumference (cm) & $96.9 \pm 14.9$ & $105.6 \pm 16.8$ \\
Diastolic Blood Pressure (mmHg) & $71.1 \pm 12.2$ & $70.9 \pm 11.2$ \\
Systolic Blood Pressure (mmHg) & $126.0 \pm 16.5$ & $119.2 \pm 15.5$ \\
Pulse (bpm) & $73.7 \pm 12.1$ & $71.5 \pm 11.36$ \\
Cholesterol (mmol/L) & $5.0 \pm 1.1$ & $4.98 \pm 1.0$ \\
Triglycerides (mmol/L) & $2.09 \pm 1.7$ & $1.55 \pm 1.06$ \\
Male Gender & $47\%$ & $50\%$ \\
Glucose (mg/dL) & $123 \pm 48$ & $89 \pm 10$ \\
HbA1c (\%) & $6.22 \pm 1.35$ & $5.33 \pm 0.35$ \\
\hline
\end{tabular}
\label{tab:describ}}
\end{table}

\subsection{Diabetes definition}

Our analysis involved 5011 participants, each assigned survey weights $w_i$ ($i=1,\dots,n=5011$). Diabetes status was defined as binary, with "1" indicating diabetes presence and "0" its absence.

The diagnostic criteria for Type 2 Diabetes, according to the American Diabetes Association (ADA) guidelines \cite{american20202}, include:
\begin{enumerate}
    \item Fasting plasma glucose (FPG) $\geq 126$ mg/dL (7.0 mmol/L), or
    \item Hemoglobin A1c (HbA1c) $\geq 6.5\%$ (48 mmol/mol), or
    \item Previous medical diagnosis of diabetes.
\end{enumerate}

Although our models incorporate FPG and HbA1c, perfect predictive accuracy remains unattainable due to individuals previously diagnosed with diabetes who currently exhibit measurements in the non-diabetic range.

\subsection{Model performance results}

We systematically evaluated seven predictive models for diabetes, assessing metrics including AUC, accuracy, recall, precision, F1 score, and cross-entropy, along with their respective operational costs. Models 1--4 progressively included more clinical variables, beginning with basic demographics and extending to physiological and metabolic markers like cholesterol and triglycerides. Model 4 significantly improved performance (AUC = 0.74; accuracy = 67\%) by incorporating metabolic markers.

Model 5 further advanced predictive performance, reaching an AUC of 0.832 and accuracy of 73.9\%, notably due to the inclusion of glycosylated hemoglobin (HbA1c). Model 7, the most comprehensive, achieved the highest performance (AUC = 0.92; accuracy = 81\%; precision = 78.6\%), benefiting from an extensive integration of clinical variables.

Our analysis reveals a clear relationship between model complexity, accuracy, and operational costs. The significant performance gains observed from Model 1 to Model 7 underscore the importance of careful variable selection, highlighting the necessity of balancing enhanced predictive accuracy with practical considerations of cost and feasibility in clinical settings.


\begin{table*}[tb!]
\caption{Performance metrics and model comparison for models 1 to 4. The table includes various clinical variables, metrics such as AUC, accuracy, recall, precision, F1 score, the cost model, cross-entropy, and the confusion matrix.}
\label{tab:performance_1_4}
\centering
{\footnotesize
\begin{tabular}{lllll}
\hline
\textbf{Variable} & \textbf{Model 1} & \textbf{Model 2} & \textbf{Model 3} & \textbf{Model 4} \\ \hline
Variable 1 & Age & Height & Age & Age \\ \hline
Variable 2 & Gender & Weight & Height & Height \\ \hline
Variable 3 & ~ & BMI & Weight & Weight \\ \hline
Variable 4 & ~ & Waist & BMI & BMI \\ \hline
Variable 5 & ~ & Diastolic Blood Pressure & Waist & Waist \\ \hline
Variable 6 & ~ & Systolic Blood Pressure & Diastolic Blood Pressure & Diastolic Blood Pressure \\ \hline
Variable 7 & ~ & Pulse & Systolic Blood Pressure & Systolic Blood Pressure \\ \hline
Variable 8 & ~ & ~ & Pulse & Pulse \\ \hline
Variable 9 & ~ & ~ & Gender & Cholesterol \\ \hline
Variable 10 & ~ & ~ & ~ & Triglycerides \\ \hline
Variable 11 & ~ & ~ & ~ & Gender \\ \hline
Cost Model & 0 & 0 & 0 & 0.5 \\ \hline
AUC           & 0.696 & 0.670 & 0.706 & 0.713 \\ \hline
Accuracy      & 0.656 & 0.668 & 0.648 & 0.648 \\ \hline
Precision     & 0.580 & 0.608 & 0.561 & 0.570 \\ \hline
Recall        & 0.632 & 0.569 & 0.713 & 0.627 \\ \hline
F$_1$--score      & 0.605 & 0.588 & 0.628 & 0.597 \\ \hline
Cross-Entropy & 0.617 & 0.618 & 0.606 & 0.604 \\ \hline
\end{tabular}}
\end{table*}

\begin{table*}[tb!]
\caption{Performance metrics and model comparison for models 5 to 7. The table continues with various clinical variables, metrics such as AUC, accuracy, recall, precision, F1 score, the cost model, cross-entropy, and the confusion matrix.}
\label{tab:performance_5_7}
\centering
{\footnotesize
\begin{tabular}{llll}
\hline
\textbf{Variable} & \textbf{Model 5} & \textbf{Model 6} & \textbf{Model 7} \\ \hline
Variable 1 & Age & Age & Age \\ \hline
Variable 2 & Height & Height & Height \\ \hline
Variable 3 & Weight & Weight & Weight \\ \hline
Variable 4 & BMI & BMI & BMI \\ \hline
Variable 5 & Waist & Waist & Waist \\ \hline
Variable 6 & Diastolic Blood Pressure & Diastolic Blood Pressure & Diastolic Blood Pressure \\ \hline
Variable 7 & Systolic Blood Pressure & Systolic Blood Pressure & Systolic Blood Pressure \\ \hline
Variable 8 & Pulse & Pulse & Pulse \\ \hline
Variable 9 & Cholesterol & Cholesterol & Cholesterol \\ \hline
Variable 10 & Triglycerides & Triglycerides & Triglycerides \\ \hline
Variable 11 & Gender & Gender & Gender \\ \hline
Variable 12 & Glycosylated Hemoglobin  & Glucose & Glucose \\ \hline
Variable 13 & ~ & ~ & Glycosylated Hemoglobin \\ \hline
Cost Model & 4.5 & 2.1 & 6.1 \\ \hline
AUC           & 0.849 & 0.826 & 0.998 \\ \hline
Accuracy      & 0.779 & 0.741 & 0.979 \\ \hline
Precision     & 0.767 & 0.784 & 0.977 \\ \hline
Recall        & 0.675 & 0.522 & 0.973 \\ \hline
F$_1$--score      & 0.718 & 0.627 & 0.975 \\ \hline
Cross-Entropy & 0.446 & 0.479 & 0.064 \\ \hline
\end{tabular}}
\end{table*}

\subsection{Uncertainity quantification}

Now, we apply the new uncertainty quantification algorithm described in Algorithm~\ref{alg:conformal}, which is based on conformal inference, to assess predictive uncertainty for the diabetes class (label “1”). Let \(\widetilde{s}(x,1)\) denote the fitted score (or predicted probability) for class “1” at the feature vector \(x\). For each observation \(i\), we define the conformal score as follows:
\[
    h_{i} \;=\; \widetilde{s}(X_{i},1) 
    \;-\; \widetilde{q}_{\alpha}(X_{i}) 
    \;+\; Q_{1-\alpha}(S, \mathcal{I}_{3}),
\]
where \(\widetilde{q}_{\alpha}(\cdot)\) is an \(\alpha\)-quantile adjustment term, and \(Q_{1-\alpha}(S, \mathcal{I}_{3})\) represents a conformal calibration constant derived via the conformal inference procedure.

To explore how uncertainty varies with patient-level covariates, we propose an \textit{additive generative model} for \(h_{i}\), using fasting plasma glucose (\(\mathrm{FPG}\)), heart rate (\(\mathrm{HR}\)), and age (\(\mathrm{AGE}\)) as predictors:

\begin{align}
h_{i} 
&= s_{\mathrm{FPG}}\bigl(\mathrm{FPG}_{i}\bigr) 
+ s_{\mathrm{HR}}\bigl(\mathrm{HR}_{i}\bigr) \notag \\
&\quad + s_{\mathrm{AGE}}\bigl(\mathrm{AGE}_{i}\bigr) 
+ \varepsilon_{i}, \quad i=1,\dots,n,
\end{align}

\noindent where \(s_{\mathrm{FPG}}, s_{\mathrm{HR}}, s_{\mathrm{AGE}}\) are smooth (cubic-spline) functions capturing the nonlinear partial effects of each covariate, and \(\varepsilon_{i}\) is an error term with zero mean.

Figure~\ref{fig:graph24} displays the estimated spline components, revealing the following:

\begin{itemize}
    \item \textbf{Highest Uncertainty Region.} The maximum value of \(h_i\) (and thus the largest predictive uncertainty) occurs for patients over 60 years old with elevated heart rate and \(\mathrm{FPG}\) values in the prediabetic range \(\,100 < \mathrm{FPG} < 125\,\mathrm{mg/dL}.\)

    \item \textbf{Lower Uncertainty Extremes.} For older individuals (60+ years) who are clearly diabetic or normoglycemic according to \(\mathrm{FPG}\), the uncertainty decreases. A similar reduction in uncertainty is observed for younger individuals with both low heart rate and low \(\mathrm{FPG}\). In these cases, the model exhibits higher discriminative power.

    \item \textbf{Clinical Interpretation.} These findings suggest that in the prediabetic zone---particularly for older adults with higher heart rates---the conformal score \(h_i\) indicates substantially greater predictive uncertainty. This, in turn, underscores the clinical need for additional testing or more frequent monitoring to confirm or exclude a diabetes diagnosis.
\end{itemize}

Overall, this analysis demonstrates how conformal inference can quantify predictive uncertainty in a patient-specific manner. From a clinical perspective, subgroups with high uncertainty (e.g., elderly patients in the prediabetic range with higher heart rates) should be considered for intensified follow-up or further diagnostic evaluations.

\begin{figure}[ht]
	\centering
	\includegraphics[width=1\linewidth,height=0.5\linewidth]{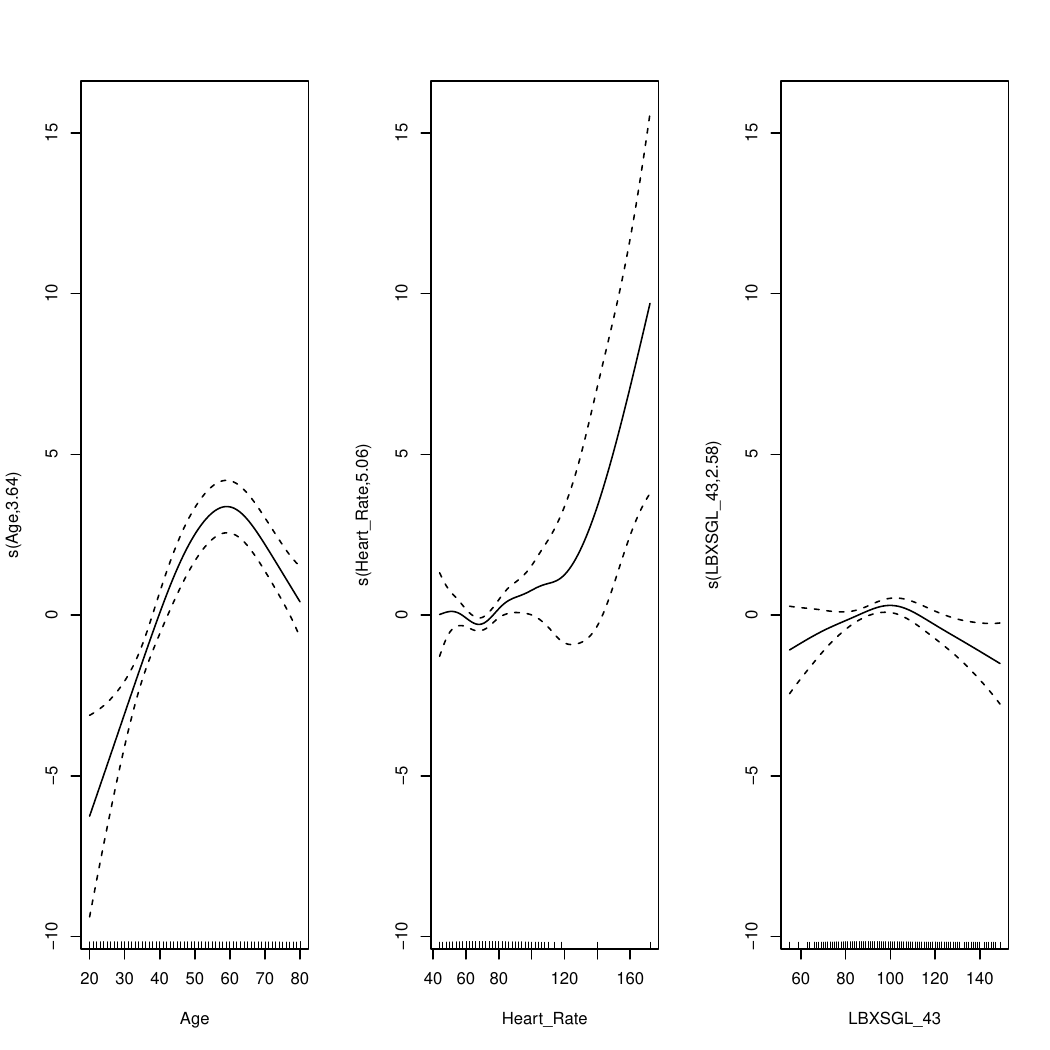} 
	\caption{Covariate effect to predict the score function $h_i$ estimated with a new conformal survey prediction algorithm. LBXSGL\_43 denotes the variable FPG.}
	\label{fig:graph24}
\end{figure}



\subsection{Experimental setup}
The dataset was partitioned into $60\%$ training, $20\%$ validation, and $20\%$ testing subsets. To ensure robust evaluation and reduce estimation bias, cross-validation was applied across all subsets, in accordance with established practices in machine learning \cite{xu2018splitting}. Model training was conducted for a maximum of $200$ epochs, with early stopping implemented using a patience parameter of $10$.

The neural network was trained using a grid search over multiple hyperparameter configurations for each scenario. Three architectures for the hidden layers were evaluated, with layer sizes set to [$32$, $16$, $8$], [$64$, $32$, $16$], and [$128$, $64$, $32$], respectively. Batch sizes were varied over ${16, 32, 64}$, and dropout rates were tested within ${0.2, 0.4, 0.6}$ to mitigate overfitting. Learning rates were explored in ${0.01, 0.001, 0.0001}$, while weight decay values were sampled from ${0, 10^{-3}, 10^{-4}}$ to regulate model complexity. The optimal hyperparameter configuration was selected via cross-validation, based on the model’s performance on the validation set. The configuration used for the experiments reported in Tables \ref{tab:performance_1_4} and \ref{tab:performance_5_7} can be found in Table \ref{tab:configurations} in the appendices.

\section{Discussion}\label{sec:discussion}


This work introduces a novel predictive framework based on neural network (NN) models capable of accommodating various complex sampling designs, including stratified designs, Bernoulli sampling, and maximum entropy sampling. A theoretical analysis provided in the supplementary materials demonstrates that the proposed regression methods achieve universal consistency under certain regularity conditions related to survey designs, analogous to the consistency obtained with independently and identically distributed (i.i.d.) data, leveraging recent advances in empirical process theory for survey data.

Through multiple simulation scenarios in a classification context, we illustrate that these methods consistently outperform classical linear logistic regression algorithms for finite-sample sizes.

In clinical applications specifically focused on predicting diabetes risk using data from NHANES 2011–2014, we evaluated models of varying complexity and associated economic costs. From a public health standpoint, our findings are significant as they quantify diabetes risk using models differing in complexity, resource requirements, and predictive efficacy. Notably, simpler models demonstrated sufficient effectiveness in certain patient subgroups, facilitating diabetes risk screening. Uncertainty quantification highlighted that age, elevated heart rate, and baseline glucose levels in the prediabetic range substantially contribute to prediction uncertainty. Despite leveraging main diagnostic criteria, perfect predictive performance is unattainable due to individuals who have normal glucose levels yet were previously diagnosed as diabetic. A model excluding laboratory measurements achieved an AUC of 0.74, comparable to traditional diabetes risk scores, by incorporating anthropometric and cardiac variables alongside basic demographic characteristics. A recent commentary \cite{mohsen2023scoping} emphasized the inconsistency of diabetes scores due to observational cohorts and limited experimental designs—a concern aligned with our introduction highlighting the reproducibility crisis. Our proposed models offer improved population-level reproducibility. Future analyses may include creating patient phenotypes based on prediction uncertainty or identifying clinically interpretable subphenotypes where models demonstrate high discriminative capacity. For patients with high prediction uncertainty or poor predictive performance, personalized monitoring strategies, alternative measurements, or the integration of longitudinal data could be essential.

From a methodological viewpoint, despite extensive research on classical regression models in survey contexts, machine learning applications remain limited. We introduce the first general NN-based framework incorporating regression and classification methods, along with a conformal prediction-based uncertainty quantification algorithm. All code and analytical scripts are publicly available on GitHub, promoting the adoption of these methods to enhance outcomes in precision public health and epidemiology through survey data. Considering that NHANES continually documents over 50,000 new clinical studies, equipping researchers and practitioners with reliable analytical tools is crucial for mitigating selection bias risks.

For future developments, several methodological extensions are recommended for the NN framework. First, expanding the approach to handle time-to-event analyses could significantly improve censored outcome predictions. Second, developing novel diagnostic tools using conditional Receiver Operating Characteristic (ROC) analyses would enhance model applicability. Third, incorporating functional data analysis techniques to manage random predictors in functional spaces could substantially benefit digital health applications, particularly when analyzing accelerometer data, minute-level measurements, or continuous glucose monitoring \cite{matabuena2023distributional}.

\bibliography{bio}
\bibliographystyle{plain}

\appendix
\section{Appendix}\label{sec:appendix}
The primary goal of this paper is to model empirical diabetes data. However, the methods introduced here are asymptotically consistent, leveraging recent progress in empirical-process theory for survey sampling. In this appendix, we sketch the essential ideas behind these developments, referring in particular to~\cite{10.1214/20-AOS1964}. 

Moreover, we connect these survey-based empirical-process techniques to established characterizations for i.i.d.\ data in the neural network literature. Specifically, existing results on metric bracketing and VC-dimension arguments—which often assume i.i.d.\ samples—can be extended to the survey setting by substituting the Horvitz--Thompson measure in place of the usual empirical measure. This substitution preserves key properties of the empirical-process framework, ensuring the validity of asymptotic consistency and normality results for neural networks trained on survey data.
\section{Theory}\label{sec:theory}

\subsection{Asymptotic neural network results}

Following~\cite{10.1214/20-AOS1964}, we introduce an empirical-process framework to derive asymptotic properties of neural networks (NN) for survey data. We adopt the super-population framework:

\begin{itemize}
\item Let $U_{N} \equiv \{1, \ldots, N\}$ be the index set for a finite population of size $N$.
\item Let $\mathcal{S}_{N}$ be the collection of all possible subsets of $U_{N}$ of size up to $N$, i.e.
\[
\mathcal{S}_{N} 
\,\equiv\, 
\bigl\{ 
   \{s_1, \ldots, s_n\} 
   :\, n \le N,\, s_i \in U_{N},\, s_i \neq s_j \text{ for } i\neq j
\bigr\}.
\]
\end{itemize}

\noindent Let $\{(Y_i, Z_i)\in\mathcal{Y}\times \mathcal{Z}\}_{i=1}^N$ be \emph{i.i.d.} samples drawn from a super-population, defined on a probability space $(\mathcal{X}, \mathcal{A}, \mathbb{P}_{(Y,Z)})$. Here
\[
Y^{(N)} \;\equiv\; (Y_1,\dots,Y_N), 
\quad 
Z^{(N)} \;\equiv\; (Z_1,\dots,Z_N).
\]

\noindent A \emph{sampling design} is a function 
\[
\mathfrak{p}:\;\mathcal{S}_{N}\times\mathcal{Z}^{\otimes N}\;\longrightarrow\;[0,1]
\]
satisfying:
\begin{enumerate}
\item[(1)] For each fixed $s\in \mathcal{S}_{N}$, the map $z^{(N)} \mapsto \mathfrak{p}(s, z^{(N)})$ is measurable.
\item[(2)] For each fixed $z^{(N)} \in \mathcal{Z}^{\otimes N}$, the map $s\mapsto \mathfrak{p}(s, z^{(N)})$ is a probability measure over $\mathcal{S}_{N}$.
\end{enumerate}

\noindent We consider the product probability space 
\[
\bigl(\,\mathcal{S}_{N}\,\times\,\mathcal{X},\;
\sigma(\mathcal{S}_{N})\,\times\,\mathcal{A},\; 
\mathbb{P}\bigr),
\]
where $\mathbb{P}$ is defined via integrals against $\mathfrak{p}$ and $\mathbb{P}_{(Y,Z)}$.  Concretely, for rectangles $s \times E$, with $s\in \mathcal{S}_N$ and $E\in \mathcal{A}$,
\[
\mathbb{P}\bigl(s\times E\bigr)
\;=\;
\int_{E} \mathfrak{p}\bigl(s, z^{(N)}(\omega)\bigr)\,d\mathbb{P}_{(Y,Z)}(\omega)
\;\equiv\;
\int_{E} \mathbb{P}_d\bigl(s,\omega\bigr)\,d\mathbb{P}_{(Y,Z)}(\omega).
\]
\noindent We use $P$ to denote the marginal law of $Y$.

Given $\bigl(Y^{(N)},Z^{(N)}\bigr)$ and a sampling design $\mathfrak{p}$, define random variables $\{\xi_{i}\}_{i=1}^{N}\subset[0,1]$ on the same space, with
\[
\pi_{i} 
\,\equiv\, 
\mathbb{E}\bigl[\xi_{i}\mid Z^{(N)}\bigr], 
\quad 
1 \,\le\, i \,\le\, N.
\]
We assume $\{\xi_{i}\}_{i=1}^{N}$ is conditionally independent of $Y^{(N)}$ given $Z^{(N)}$.  A typical choice is $\xi_{i}=\mathbf{1}_{\{i\in s\}}$ for $s\sim \mathfrak{p}$, in which case 
\(
\pi_{i}(Z^{(N)})
=\sum_{s:i\in s}\,\mathfrak{p}(s,Z^{(N)}).
\)
Then $\pi_{i}$ is the \emph{first-order inclusion probability} and $\pi_{ij}$ is the \emph{second-order inclusion probability} given $Z^{(N)}$.

\subsubsection{Horvitz--Thompson empirical measure and process}

Define the \emph{Horvitz--Thompson} (HT) empirical measure for a function $f\in \mathcal{F}$:
\[
\mathbb{P}_{N}^{\pi}(f)
\;\equiv\; 
\frac{1}{N}\,\sum_{i=1}^{N}\,\frac{\xi_{i}}{\pi_{i}}\;f(Y_i),
\]
and its associated empirical process
\[
\mathbb{G}_{N}^{\pi}(f)
\;\equiv\; 
\sqrt{N}\,\bigl(\mathbb{P}_{N}^{\pi}-P\bigr)(f), 
\quad 
f\in \mathcal{F}.
\]
Here $P(f)=\mathbb{E}[f(Y)]$ is the super-population expectation.  The term ``Horvitz--Thompson'' stems from~\cite{horvitz1952generalization}, where $\mathbb{P}_{N}^{\pi}(Y)$ is shown to be an unbiased estimator of the population mean $P(Y)$.  The usual empirical measure $\mathbb{P}_{N}$ (i.e.\ $\xi_{i}/\pi_{i}\equiv 1$) and its empirical process $\mathbb{G}_{N}$ arise as a special case.

\begin{assumption}[Sampling design conditions]
\label{assump:A}
We consider the following conditions on the sampling design $\mathfrak{p}$:
\begin{enumerate}[label=\textbf{(A\arabic*)}, leftmargin=1.55cm]
\item \label{A1} 
\(\min_{1 \le i \le N}\,\pi_{i} \;\ge\; \pi_{0}>0.\)

\item \label{A2-LLN} \(\dfrac{1}{N}\,\sum_{i=1}^{N}\Bigl(\frac{\xi_{i}}{\pi_{i}}-1\Bigr) \;=\; \mathfrak{o}_{\mathbf{P}}(1)\).

\item \label{A2-CLT} \(\dfrac{1}{\sqrt{N}}\,\sum_{i=1}^{N}\Bigl(\frac{\xi_{i}}{\pi_{i}}-1\Bigr) \;=\; \mathcal{O}_{\mathbf{P}}(1)\).
\end{enumerate}
\end{assumption}

\noindent Above, \(\mathfrak{o}_{\mathbf{P}}(1)\) means the quantity converges to 0 in probability, and \(\mathcal{O}_{\mathbf{P}}(1)\) means it is stochastically bounded.

\subsubsection{Key theorems from \cite{10.1214/20-AOS1964}}

Below are classical results (Glivenko--Cantelli and Donsker theorems) specialized to the HT empirical process under Assumption~\ref{assump:A}.

\begin{theorem}[Glivenko--Cantelli]
\label{thm:glivenko-cantelli}
\cite{10.1214/20-AOS1964}. 
Assume \textnormal{\ref{A1}} and \textnormal{\ref{A2-LLN}} hold.  If $\mathcal{F}$ is a $P$-Glivenko--Cantelli class, then
\[
\sup_{f \in \mathcal{F}} 
\Bigl|\,\bigl(\mathbb{P}_{N}^{\pi}-P\bigr)(f)\Bigr|
\;=\;
o_{P}(1).
\]
\end{theorem}

\begin{theorem}[Donsker]
\label{thm:donsker}
\cite{10.1214/20-AOS1964}. 
Assume \textnormal{\ref{A1}} and \textnormal{\ref{A2-CLT}} hold, and further suppose:
\begin{enumerate}[label=\textbf{(D\arabic*)}, leftmargin=1.2cm]
\item \(\mathbb{G}_{N}^{\pi}\) converges finite-dimensionally to a tight Gaussian process \(\mathbb{G}^{\pi}\).
\item \(\mathcal{F}\) is $P$-Donsker.
\end{enumerate}
Then
\[
\mathbb{G}_{N}^{\pi} 
\;\rightsquigarrow\; 
\mathbb{G}^{\pi}
\quad 
\text{in } \ell^{\infty}(\mathcal{F}),
\]
where ``\(\rightsquigarrow\)'' denotes weak convergence in the supremum norm.
\end{theorem}

\subsection{Neural networks for survey data: Asymptotics}

We now leverage these empirical-process results to study the asymptotic behavior of neural networks (NNs) trained under survey sampling.

\begin{theorem}[Consistency and asymptotic normality]
\label{thm:NN-consistency-AN}
Consider a neural network architecture with ReLU activation.  Let the loss function \(\ell(\cdot,\cdot)\) be one of: 
\begin{itemize}
\item weighted cross-entropy, 
\item pinball loss, or 
\item mean squared error (MSE).
\end{itemize}
Under Assumptions~\ref{A1}--\ref{A2-CLT} (or stronger versions if needed), the resulting NN parameter estimators are consistent.  Moreover, these estimators are asymptotically normal, by virtue of the invariance principle for the (Donsker) empirical process defined by the chosen loss.
\end{theorem}
\begin{proof}
  The scheme proof uses:
\begin{enumerate}
\item[(i)] Finite VC-dimension arguments for ReLU networks~\cite{bartlett2019nearly}.
\item[(ii)] Bracketing-entropy methods for the functional classes~\cite{rynkiewicz2019asymptotic}.
\item[(iii)] The Horvitz--Thompson empirical measure replacing the usual empirical measure.
\end{enumerate}
These steps collectively imply a Donsker property for the underlying loss functions, yielding asymptotic normality and consistency.
\end{proof}

\begin{theorem}[Asymptotic coverage of uncertainty sets]
\label{thm:coverage}
Let \(\alpha\in(0,1)\) be a confidence level.  Consider an uncertainty-quantification procedure (e.g.\ a conformal method) that produces prediction sets \(\widetilde{C}^{\alpha}(X_{n+1};\mathcal{D}_n)\).  Although exact finite-sample coverage
\[
\mathbb{P}\!\Bigl(
  Y_{n+1}\,\in\, \widetilde{C}^{\alpha}(X_{n+1};\mathcal{D}_{n})
\Bigr)
\;\ge\; 
1-\alpha
\]
may fail in general, the procedure is \emph{asymptotically consistent} for both marginal and conditional coverage.  In particular,
\[
\lim_{n\to \infty} 
\Bigl|\,
  (1-\alpha) 
  \;-\; 
  \mathbb{P}\!\Bigl(
    Y_{n+1}\in \widetilde{C}^{\alpha}(X_{n+1};\mathcal{D}_{n})
  \Bigr)
\Bigr|
\;=\;
0
\quad
\text{in probability.}
\]
\end{theorem}
\begin{proof}
A-weighted variant of the proof of \cite{cauchois2021knowing} but the same steps. 
The main idea relies on:
\begin{enumerate}
\item[(i)] Mean-square consistency of the fitted (NN) score functions and empirical quantile estimators,
\item[(ii)] Asymptotic properties of the Horvitz--Thompson empirical measure (Theorems~\ref{thm:glivenko-cantelli}--\ref{thm:donsker}).
\end{enumerate}
Together, these imply that for large $n$, the coverage probability of $\widetilde{C}^{\alpha}(\cdot;\mathcal{D}_n)$ converges to $1-\alpha$ under both marginal and conditional viewpoints.
\end{proof}

\begin{remark}[Extensions]
Similar arguments hold for the pinball loss or MSE if the bracketing entropy conditions are satisfied, as in \cite{rynkiewicz2019asymptotic}.  One only needs to replace the usual empirical measure with the Horvitz--Thompson measure in all limiting arguments.
\end{remark}

\begin{table*}[tb!]
\centering
\caption{Best model configuration for the experiments reported in Tables \ref{tab:performance_1_4} and \ref{tab:performance_5_7}.}
\label{tab:configurations}
\scalebox{0.65}{
\scriptsize
\begin{tabular}{@{}llllllll@{}}
\hline
\textbf{Parameter} & \textbf{Model 1} & \textbf{Model 2} & \textbf{Model 3} & \textbf{Model 4} & \textbf{Model 5} & \textbf{Model 6} & \textbf{Model 7} \\ \hline
Hidden layers & [32, 16, 8] & [64, 32, 16] & [128, 64, 32] & [128, 64, 32] & [64, 32, 16] & [64, 32, 16] & [64, 32, 16] \\ \hline
Batch size    & 32           & 64           & 32            & 32            & 16            & 16            & 16            \\ \hline
Dropout       & 0.0          & 0.2          & 0.5           & 0.5           & 0.0           & 0.5           & 0.0           \\ \hline
Learning rate & 0.01         & 0.001        & 0.0001        & 0.0001        & 0.01          & 0.0001        & 0.01          \\ \hline
Weight decay  & 0.001        & 0.0001       & 0.001         & 0.0001        & 0.0           & 0.001         & 0.0           \\ \hline
\end{tabular}
}
\end{table*}

\subsection{MLP configuration}

Table~\ref{tab:configurations} presents the optimal model configurations identified for the experiments reported in Tables~\ref{tab:performance_1_4} and~\ref{tab:performance_5_7}, conducted as part of the NHANES Diabetes Case Study. The table summarizes the key hyperparameters that yielded the best performance across the experimental runs, including the architecture of the hidden layers, batch size, dropout rate, learning rate, and weight decay. These configurations were selected based on cross-validation results and reflect the most effective settings for optimizing model generalization and stability within the studied context.

\end{document}